\documentclass[pra, 10pt, twocolumn, letter, floatfix]{revtex4}
\usepackage{amsmath}
\usepackage{graphicx}
\usepackage{rotating}
\allowdisplaybreaks
\usepackage{setspace}

\begin{document}

\title{A correlator product state study of molecular magnetism\\ in the giant {Keplerate} Mo$_{72}$Fe$_{30}$}

\author{Eric Neuscamman$\mathrm{{}^\ast}$ and Garnet Kin-Lic Chan$\mathrm{{}^\dagger}$}
\affiliation{$\mathrm{{}^\ast}$Department of Chemistry, University of California, Berkeley, California 94720, USA\\
             $\mathrm{{}^\dagger}$Department of Chemistry, Princeton University, Princeton, NJ 08544, USA}


\date{\today}

\begin{abstract}
We have studied the properties of the giant Keplerate molecular magnet $\mathrm{Mo_{72}Fe_{30}}$, as a function of
 applied magnetic field, using the correlator product state (CPS) tensor network ansatz.
 The magnet is modeled with an $S=5/2$ antiferromagnetic Heisenberg Hamiltonian on the 30-site
icosidodecahedron lattice, a model for which exact diagonalization is infeasible.
The CPS ansatz produces significant improvements in variational energies relative to previous
studies using the density matrix renormalization group, a result of its superior ability to
handle strong correlation in two dimensional spin systems.
The CPS results reaffirm that the ground state energies adhere qualitatively to the parabolic
progression of the rotational band model (RBM), but show important deviations near 1/3 of the
saturation field. These deviations predict anomalous behavior in the differential
magnetization and heat capacity that cannot be explained by the RBM alone.
Finally, we show that these energetic deviations originate from a qualitative change
in the ground state that resembles a finite size analogue of a phase transition.
\end{abstract}

\maketitle


%
%
%

\section{Introduction}
\label{sec:introduction}



\begin{figure}[t]
\centering
\includegraphics[width=8.5cm,angle=0]{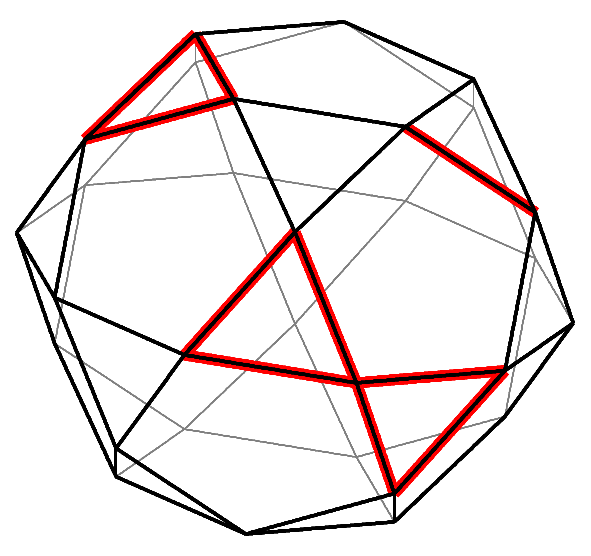}
\caption{The giant Keplerate $\mathrm{Mo_{72}Fe_{30}}$ molecular magnet is shaped
         like an icosidodecahedron, with the Fe atoms positioned on the vertices
         and the $\mathrm{-O-Mo-O-}$ bridges along the edges.
         Three example correlators have been shaded in red:  a two-site nearest neighbor,
         a three-site triangle, and a five-site bow tie.
        }
\label{fig:icosidodecahedron}
\end{figure}

Molecular magnets are classic examples of chemical systems containing a large number
of localized, strongly correlated electrons.
Their study has been motivated both from potential applications in
storage and quantum computing, as well as by the fundamental challenges associated
with their chemical synthesis and their physical magnetic properties
\cite{KAHN:1993:mol_magnetism,
      Gatteschi:1994:mol_magnetism,
      BLUNDELL:2004:org_mol_mag,
      GATTESCHI:2006:mol_magnets,
      BLUNDELL:2007:mol_magnets,
      SCHNACK:2010:frust_mol_mag}.
In recent years, using polyoxometalate chemistry \cite{MULLER:2001:polyoxometalate}, some very large
molecular magnets have been synthesized \cite{MULLER:1999:linkable_units,MULLER:1999:keplerate}.
These so-called giant Keplerate magnets earn their name from the geometric arrangement of
the ions, which lie at the vertices of regular solids. 
The largest such magnet made to date is based on the icosidodecahedron, and consists of corner sharing
triangles arranged around pentagons (see Figure \ref{fig:icosidodecahedron}).
The metal species can be varied, and magnets including V, Cr, and Fe ions have been made,
although the Fe based Keplerate magnet has been the most studied so far
\cite{FU:2010:neutron_scatter,
      LUBAN:2006:neutron_scatter,
      MILA:2008:kagome_sphere,
      MULLER:2001:giant_keplerate,
      SCHNACK:2001:rot_band_model,
      SCHNACK:2003:mo72fe30_dmrg,
      SCHNACK:2005:frust_effects,
      SCHRODER:2005:diff_susc,
      SCHRODER:2008:random_exchange}.
The corner sharing triangle geometry leads to magnetic frustration and unusual magnetic properties
\cite{SCHNACK:2010:frust_mol_mag} which are of interest in this work.


\begin{figure}[b]
\centering
\includegraphics[width=8.5cm,angle=0]{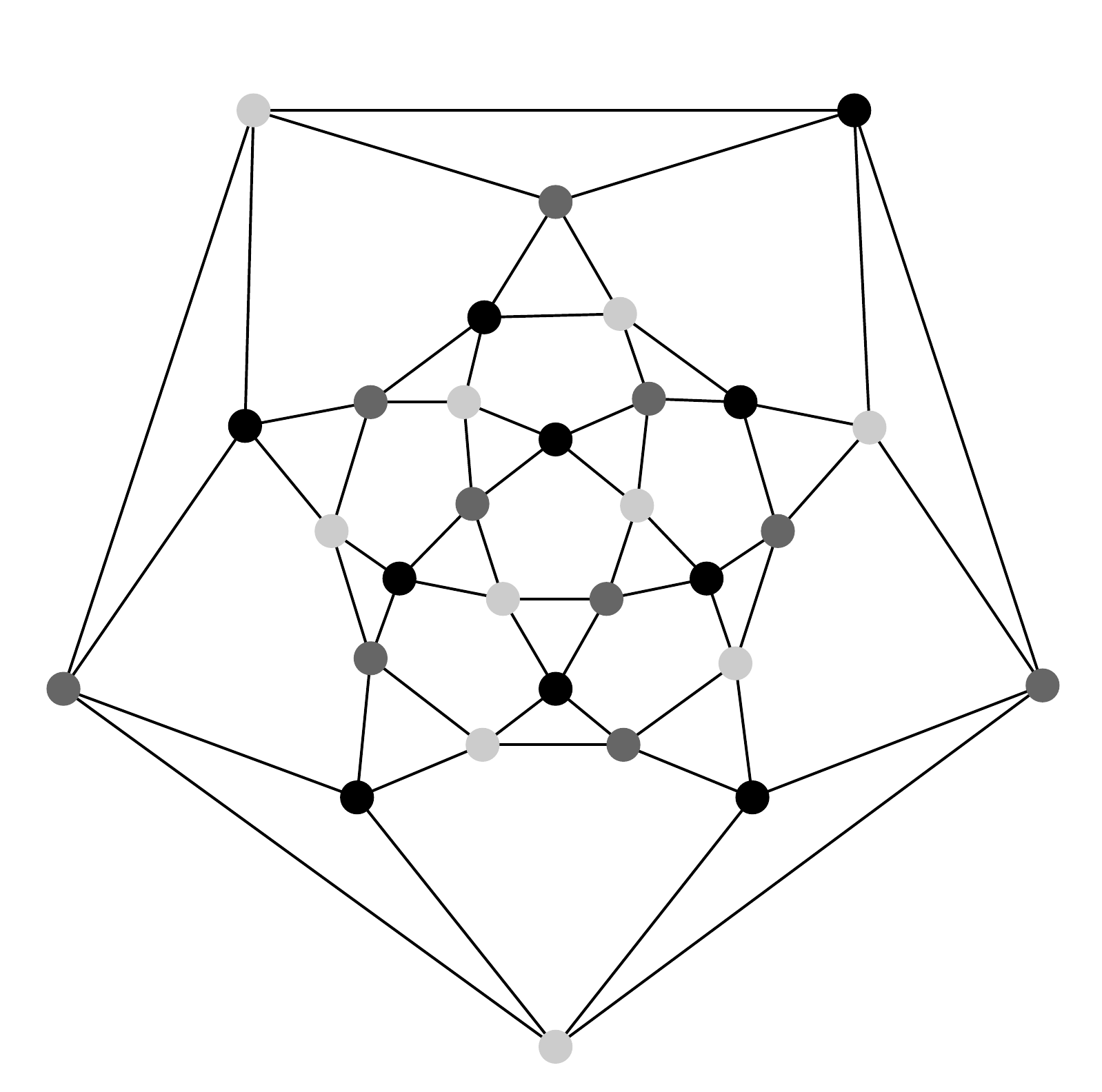}
\caption{The icosidodecahedron lattice flattened to a planar graph.
         The vertices are shown under the particular three-coloring that was
         used to generate the initial guess for our CPS wavefunction optimization.
        }
\label{fig:coloring}
\end{figure}

The theoretical description of magnetism in the Keplerate magnets is extremely challenging.
The basic reason is the size of the Hilbert space associated with the magnetic centers.
In the case of the Fe$_{30}$-Keplerate, each Fe center is a 3+ ion with 5 unpaired spins in
a near perfect octahedral coordination, and we can view each center as effectively an $S=5/2$ spin
\cite{MILA:2008:kagome_sphere}. 
Arranging the spins on the vertices of the icosidodecahedron
(see Figures \ref{fig:icosidodecahedron} and \ref{fig:coloring}),
we model their interactions using the Heisenberg Hamiltonian,
\begin{align}
\label{eqn:heisenberg_ham}
H = J \sum_{\langle ij \rangle} S_i \cdot S_j
\end{align}
where $\langle ij \rangle$ represents a summation over nearest neighbors.
Since there are 30 $S=5/2$ spins, the corresponding Hilbert space is of dimension
$6^{30}$, or roughly $10^{23}$, a mole of quantum states!
This is far too large to treat using the exact diagonalization methods that are usually
employed for molecular magnets \cite{SCHNACK:2010:ed_symmetry}. 


In this work, we use a variational methodology based on {\it correlator product states} (CPS)
\cite{CHANGLANI:2009:cps,NEUSCAMMAN:2011:pcps}, in conjunction with the Heisenberg
Hamiltonian, to model the low-lying states of the Fe$_{30}$-Keplerate magnet.
Correlator product states, also known as entangled plaquette states
\cite{MEZZACAPO:2009:entangled_plaquettes,MEZZACAPO:2010:entangled_plaquettes}
or complete graph tensor networks \cite{VERSTRAETE:2010:cgtn}, provide a simple
approximation to the quantum wavefunction amplitude for a large number of spins as a product of amplitudes of
smaller overlapping subsets of spins. The term correlator refers to the amplitudes on the subsets of spins.
The CPS approximation derives from an attempt to generalize the density matrix renormalization group
(DMRG) \cite{WHITE:1992:dmrg}, a powerful method for strongly correlated electrons that
has been applied both to realistic quantum chemical problems
\cite{WHITE:1992:qc_dmrg,
      CHAN:2002:dmrg,
      HACHMANN:2006:dmrg,
      GHOSH:2008:dmrg,
      CHAN:2010:dmrg_ct,
      SHARMA:2011:qc_dmrg_review}
as well as many model condensed matter Hamiltonians
\cite{WHITE:2012:2d_dmrg,SCHOLLWOCK:2005:dmrg,HALLBERG:2006:dmrg}.
Exler and Schnack previously used the DMRG to study the Fe$_{30}$-Keplerate magnet
\cite{SCHNACK:2003:mo72fe30_dmrg}, providing a qualitative demonstration of the existence of
a quantum rotational band.
However, the DMRG has difficulty in accurately treating large systems where correlations
are not ordered in a one-dimensional fashion. Unlike the DMRG, the CPS is not biased towards one-dimensional correlations,
and thus in principle can be an efficient ansatz for the correlations present in the Keplerate magnets.
Here we will compare our CPS calculations not only to the available experimental measurements, but also to the
 earlier theoretical DMRG work of Exler and Schnack,  demonstrating the improved ability of the CPS to describe correlations in general
 systems.

The structure of our study is as follows.
We first give an overview of the theoretical and experimental results for magnetism in the
giant Keplerate magnets and discuss, in particular, features related to magnetic frustration in the
icosidodecahedron (Section \ref{sec:keplerate_magnetism}).
We then describe the general theory behind the CPS wavefunction (Section \ref{sec:theory}),
how it is optimized via variational Monte Carlo (Section \ref{sec:vmc}), and the specific
form of the wavefunction we use in this work (Section \ref{sec:wfn_details}).
We next present the results of our calculations in light of experimental
and earlier theoretical work on the magnet. In particular, we present
total energies (Section \ref{sec:total_energies}),
low temperature properties (Section \ref{sec:properties}),
spin correlations (Section \ref{sec:spin_correlations}),
and an analysis of possible phase transition behavior (Section \ref{sec:phase_transition}).
Finally, we conclude with some perspectives for further work on the Keplerate systems,
and ways to generalize the CPS approach to other complex molecular systems (Section \ref{sec:conclusions}).

\section{Magnetism in the giant Keplerates}
\label{sec:keplerate_magnetism}

%
%
%

Keplerate systems are interesting from the viewpoint
of quantum magnetism due to the presence of frustration effects \cite{SCHNACK:2010:frust_mol_mag}.
One way to define a frustrated magnet is one where the classical Ising model, whose spins only
assume up ($u$) and down ($d$) orientations, has a large degeneracy. 
This is the case for triangles, where the $uud$, $udu$, and $duu$ configurations are
all degenerate.  In the classical Heisenberg model, where spins can point in any orientation, 
the spin triangle has a continuous manifold of degenerate
ground states. In these states, the three spins are coplanar and rotated $120^\circ$ from each other, and it
is the orientation of the plane that creates the continuous degeneracy.

While a single spin triangle already shows some frustration effects, such effects become even
more pronounced in the case of {\it corner sharing} triangles \cite{SCHNACK:2010:frust_mol_mag}.
This is the motif underlying the icosidodecahedron, whose surface consists of corner sharing
triangles arranged around pentagons.
In fact, the icosidodecahedron is the largest member of a family of Platonic solids, which also
includes the cuboctahedron and the truncated tetrahedron, whose surfaces are built from corner
sharing triangles.
The quantum Ising model on these lattices is highly frustrated.
These zero dimensional  systems are especially important as they exist as finite size surrogates for
their bulk planar counterparts, such as the two dimensional Kagome lattice
\cite{WHITE:2011:kagome_dmrg}, which are believed to underlie exotic magnetism in solids.
Because of their small size, the Platonic solid models allow the effects of corner sharing triangle
frustration to be studied in an experimentally realizable system that is also accessible to many theoretical approaches.

We now give a brief overview of some of the interesting properties that can arise from spin frustration
in corner sharing triangle systems. One class of frustration effects is the presence of anomalies that occur at applied
magnetic field strengths close to $1/3$ of the saturation field $B_{sat}$. (The saturation
field is the field strength above which the ground-state has all spins aligned with the field).
It has been observed both experimentally
\cite{SCHRODER:2005:diff_susc,SCHRODER:2008:random_exchange} and theoretically
\cite{SCHRODER:2005:diff_susc,SCHRODER:2008:random_exchange,ZHITOMIRSKY:2002:kagome} that the
differential susceptibility $dM/dB$ (the rate of change of the total system magnetization with
respect to field strength) displays a depression near field strengths of $B_{sat}/3$.
A rough understanding of this is that near $B_{sat}/3$, the magnetically stiff
$uud$ states of the spin triangles become energetically competitive with the usual ground-state,
but this alone only gives a qualitative accounting of the experimental data.
In Keplerate systems, a more quantitative match \cite{SCHRODER:2008:random_exchange} to the
observed $dM/dB$ depression was achieved under the assumption of random variations in the spin
couplings within the classical Heisenberg model.
Note that the $B_{sat}/3$ anomaly does not only appear in the differential susceptibility, but
also shows up, for example, in zero temperature magnetization predictions
\cite{MILA:2008:kagome_sphere}, in the heat capacity \cite{ZHITOMIRSKY:2002:kagome}, and as a
phase transition in the classical Heisenberg model on the Kagome lattice
\cite{ZHITOMIRSKY:2002:kagome,GVOZDIKOVA:2011:phase_diag}.

Another interesting aspect of frustrated spin systems is the possibility of unusually
low-lying singlet excited states. Although these states will not be treated in this study, they have attracted a great
deal of interest \cite{MILA:2008:kagome_sphere,SCHNACK:2005:frust_effects,SCHNACK:2009:exact_diag}
and are implicated as a means to explain puzzling experimental neutron scattering data
\cite{LUBAN:2006:neutron_scatter,FU:2010:neutron_scatter} in the giant Keplerate magnets.

To a first approximation (although  see Ref.\ \cite{SCHRODER:2008:random_exchange})
 magnetism in these systems can be described by an isotropic
Heisenberg  model, with the M ion coupled antiferromagnetically via the Mo-O bridges.
The M ions are believed to lie in near perfect $O_h$ coordination with the oxygens, and the V, Cr, and Fe giant
Keplerates can be thought of as $S=1/2$, 3/2, and 5/2 spin centers.
Most experimental work has focused on characterizing the $S=5/2$ Fe system, and it is the corresponding
$S=5/2$ Heisenberg model on the icosidodecahedron to which we apply the CPS wavefunction.

The essential problem in studying the icosidodecahedron Heisenberg model is the very large Hilbert
space that needs to be considered, which is $2^{30}$, $4^{30}$, $6^{30}$ for the V, Cr, Fe species.
Exact (full) diagonalization of the Heisenberg Hamiltonian has been carried out in the $S=1/2$ case of the V$_{30}$ magnet
but is impossible for the other magnets.
Nonetheless, many of the qualitative features of these systems appear to be well described by a rather simple
model known as the quantum rotational band model.
The Keplerate magnets are tripartite (see Figure \ref{fig:coloring}), and we can consider therefore a family of
spins living on the A, B, C sub-lattices.
The quantum rotational band model (RBM) \cite{SCHNACK:2001:rot_band_model} asserts that the energies of the states
can be modeled as arising from the couplings of {\it total spins on the A and B and C lattice as an effective triangle}
and is given by the Hamiltonian
\begin{align}
H_{band} = J \frac{D}{N} \left [ \vec{S}^2 - \gamma \left( \vec{S}^2_A + \vec{S}^2_B + \vec{S}^2_C \right) \right],
\end{align}
where $N$ is the number of spins, $D$ and $\gamma$ are free parameters, $\vec{S}$ is the net lattice spin, and
$\vec{S}_A$, $\vec{S}_B$, and $\vec{S}_C$ are the net spins on each sublattice.
The eigenstates of this Hamiltonian have energies
\begin{align}
& E(S, S_A, S_B, S_C) \notag \\
& ~ = J \frac{D}{N} \left[ S(S+1) - \gamma \left( \sum_{ Q \in \{ A, B, C \} } S_Q(S_Q+1) \right) \right]
\label{eqn:rbm_energy}
\end{align}
with degeneracies given by the number of ways a given total spin $S$ can be made up from the sublattice spins via spin
coupling rules.
As will be discussed in the results below, the RBM has been successful at qualitatively reproducing experimental
magnetizations \cite{SCHNACK:2001:rot_band_model} as well as the total energies of some higher level theoretical
treatments \cite{SCHNACK:2003:mo72fe30_dmrg}.
However, as we will demonstrate in the case of differential susceptibility, it fails to predict
the peculiar properties of the $\mathrm{Mo_{72}Fe_{30}}$ Keplerate magnet related to the $B_{sat}/3$ anomaly.
For these effects we need to consider all the spin degrees of freedom, for which we need explicit approximations
for the quantum wavefunction.


\section{Correlator product states}
\label{sec:cps}

\subsection{General Theory}
\label{sec:theory}

Consider a set of $k$ spins $s_1 \ldots s_k$.
In an $S=5/2$ system, such as in the Fe$_{30}$-Keplerate magnet, each $s$ varies over
the 6 $m_s$ levels of each iron center.
The quantum wavefunction  written in the complete spin Hilbert space is
\begin{align}
\label{eqn:general_wf}
|\Psi\rangle &= \sum_{s_1 s_2 ... s_k} \Psi_{s_1 s_2 ... s_k} | s_1 s_2 ... s_k \rangle \\
             &= \sum_{\mathbf{s}} \Psi_{\mathbf{s}} |\mathbf{s}\rangle \notag
\end{align}
where $\mathbf{s}$ denotes the vector of spin configurations $s_1 s_2 \ldots s_k$.

The amplitude $\Psi_{s_1 s_2 ... s_k}$ is infeasible to obtain exactly for a system
as large as the Fe$_{30}$-Keplerate magnet. Correlator product states provide an approximation
for the full amplitude in terms of simpler objects known as correlators.
In spin systems, a correlator defines a set of amplitudes over a subset (domain) of the spin sites.
For example, a correlator on sites $i, j$ defines a set of amplitudes $c_{s_i s_j}$.
 Correlators can be constructed to act on an arbitrary number of sites
 (see Figure \ref{fig:icosidodecahedron}).
 Such a general correlator is  written as $c_{\mathbf{s}_\lambda}$ where $\mathbf{s}_\lambda$
denotes the spin configuration of the subset of sites $\lambda$.
To obtain the CPS, we approximate the wavefunction amplitudes $\Psi_{\mathbf{s}}$ in Eq.\ (\ref{eqn:general_wf})
as a product of correlator amplitudes over the different subsets of sites $\lambda$,
\begin{align}
\label{eqn:ansatz}
\Psi_\mathbf{s} = \prod_\lambda c_{\mathbf{s}_\lambda}
\end{align}
Note that the domains $\lambda$ of the different correlators will usually contain
overlapping sites.
For example,  a CPS wavefunction for a one-dimensional arrangement of spins with ``nearest neighbor'' correlators,
would be written as
\begin{align}
\label{eqn:cps1d_operator}
\Psi_{s_1s_2 \ldots s_k} = c_{s_1 s_2} c_{s_2 s_3} \ldots c_{s_{k-1}s_k}
\end{align}
By using correlators that cover increasingly larger numbers of sites, we can make the CPS approximation
arbitrarily exact.

\subsection{Monte Carlo Optimization}
\label{sec:vmc}

We use the variational Monte Carlo algorithm to optimize
the CPS wavefunction to obtain approximate ground-states of the Keplerate magnet.
(We have shown elsewhere that the CPS wavefunction can also be used
with non-stochastic algorithms \cite{NEUSCAMMAN:2011:pcps}, although these are not employed here).
In variational Monte Carlo, the energy is written as
\begin{align}
\label{eqn:mc_energy}
E &= \frac{\langle\Psi|H|\Psi\rangle}{\langle\Psi|\Psi\rangle} \\
  &=  \sum_{\mathbf{s}} \frac{|\Psi_\mathbf{s}|^2}{\langle\Psi|\Psi\rangle} E_{\mathrm{L}}(\mathbf{s}) \notag 
\end{align}
where the local energy $E_{\mathrm{L}}(\mathbf{s})$ is defined by
\begin{align}
\label{eqn:mc_local_energy}
E_{\mathrm{L}}(\mathbf{s}) = & \sum_{\mathbf{s}'} \frac{\Psi_{\mathbf{s}'}}{\Psi_{\mathbf{s}}}
                                    \langle \mathbf{s} | \hat{H} | \mathbf{s}'\rangle.
\end{align}
As long as $\Psi_{\mathbf{s}}$ can be evaluated efficiently, which is the case for the CPS
wavefunctions, a Markov chain can be used to sample the probability distribution
$|\Psi_\mathbf{s}|^2/\langle \Psi|\Psi\rangle$ and efficiently compute the overall
energy as an average of the sampled local energies.
The energy is then variationally minimized using stochastic estimates for the
gradient with respect to the correlator amplitudes. Note that it is easy to constrain
the Monte Carlo sampling over ${\mathbf{s}}$, for example 
in  Eq.\ (\ref{eqn:mc_energy}),  to only those configurations with a given value of $S_z$,
 and this allows us to obtain approximate ground-states in different $S_z$ sectors. 
Once the wavefunctions are obtained, expectation values for various correlation functions
can also be readily computed by Monte Carlo sampling.

\subsection{Wavefunction and Optimization  Details}
\label{sec:wfn_details}

To study the Fe$_{30}$-Keplerate magnet we used a CPS in the form of Eq.\ (\ref{eqn:ansatz}) with bow tie shaped
correlators.
There are 30 different bow ties in all, each defined by choosing one site and all of its nearest neighbors
(see Figure \ref{fig:icosidodecahedron} for an example).
After discovering that randomly chosen correlator amplitudes were not effective as initial guesses for
the variational optimization, we chose to use as our guess a relatively simple state similar to the
classical ground state.
To be precise, our initial guess for $S_z=0$ was chosen to be a spin-coherent state
\cite{SHIBATA:1976:scs} which can be exactly represented by a CPS. In a spin-coherent state, the wavefunction amplitude factorizes into a product
of amplitudes on individual sites $\Psi_{s_1s_2 \ldots s_k} = c_{s_1}c_{s_2} \ldots c_{s_k}$, and each site amplitude
$c_{s_i}$ defines a direction for the spin on the site. Here we chose
the
  rotation angles for each site to be the
classical ground state's spin direction for that site's sublattice (the sublattices were assigned based
on the coloring shown in Figure \ref{fig:coloring}).
Starting from this guess, we optimized the wavefunction's energy under $S_z=0$ projection, and then used
the resulting wavefunction as an initial guess for the $S_z=1$ sector.
In this fashion we worked our way up the magnetization ladder, obtaining a wavefunction for each $S_z$
sector.
To help ensure convergence, we then worked backwards, using the
$S_z=74$ solution for the $S_z=73$ guess
and re-optimizing, retaining whichever wavefunction gave the lowest energy before moving down to the next
$S_z$ sector.
This sweeping procedure was especially helpful for resolving the minimum energies for $S_z\le30$.

\begin{figure}[t]
\centering
\includegraphics[width=8.5cm,angle=0]{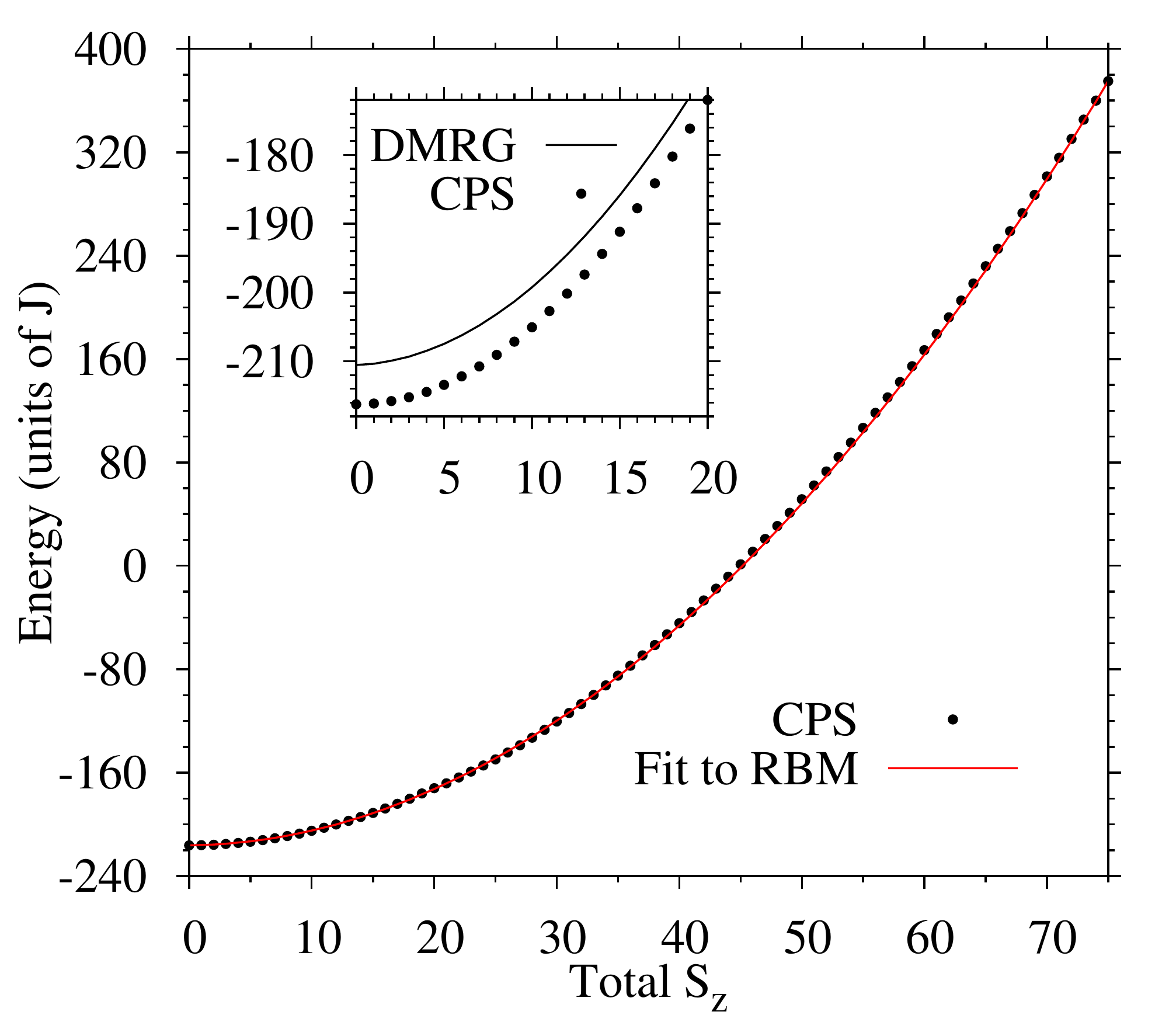}
\caption{Ground state energies for the $S=5/2$ Heisenberg model on the icosidodecahedron for different
         total $S_z$ sectors of the Hilbert space.
         In the main panel the CPS wavefunction's energies are shown along with the corresponding
         fit to the RBM form, Eq.\ (\ref{eqn:rbm_energy}).
         In the inset the CPS energies are compared to the RBM produced by fitting DMRG energies
         \cite{SCHNACK:2003:mo72fe30_dmrg}.
         See Section \ref{sec:total_energies}.
        }
\label{fig:total_energy}
\end{figure}

\section{Results}
\label{sec:results}

\subsection{Total energies}
\label{sec:total_energies}



\begin{figure}[t]
\centering
\includegraphics[width=8.5cm,angle=0]{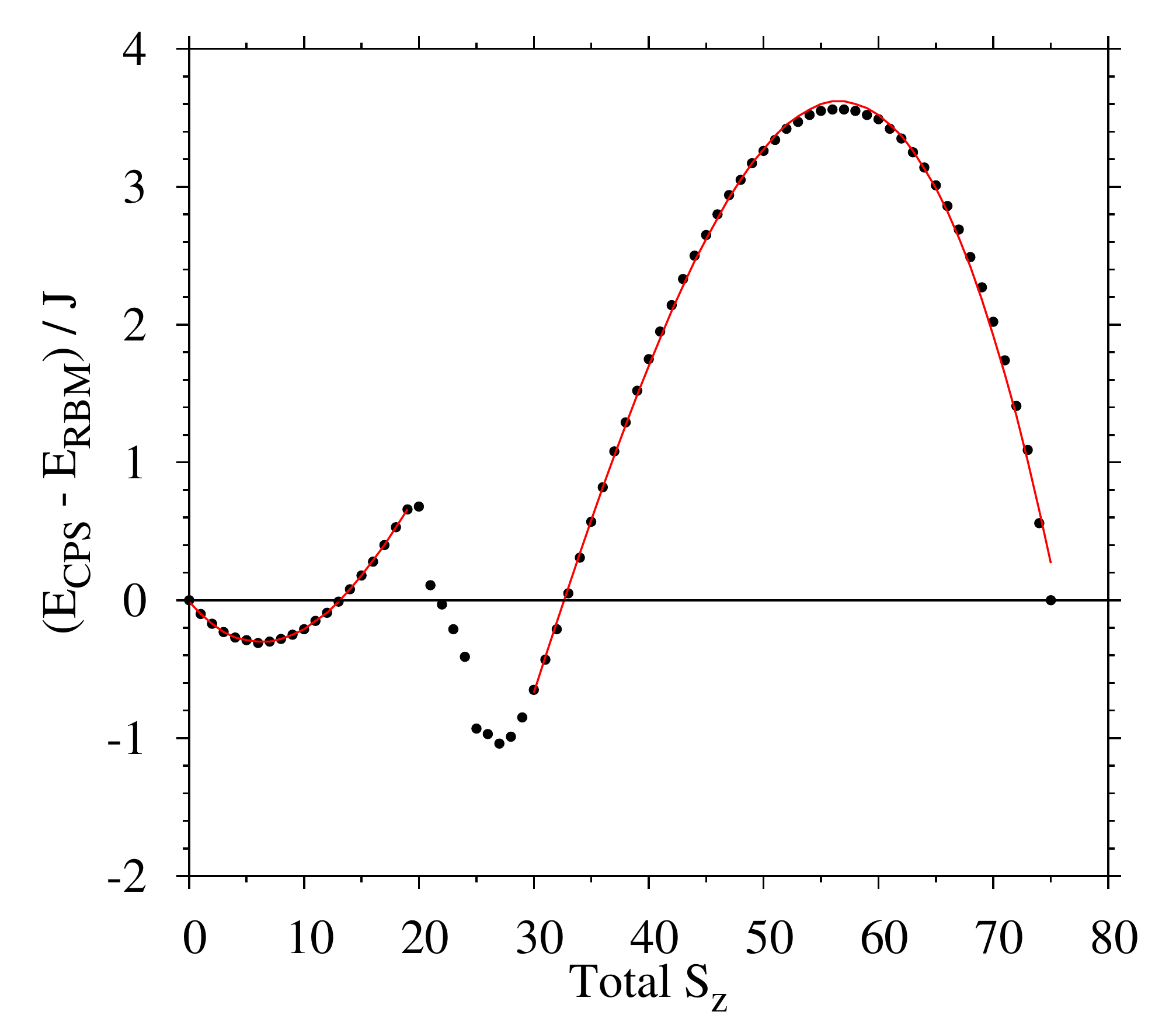}
\caption{Deviation of the CPS energies from the fit to the RBM.
         Circles represent the raw deviations, while lines represent cubic fits in the ranges
         $S_z \in [0,19]$ and $S_z \in [30,75]$.
         See Section \ref{sec:total_energies}.
        }
\label{fig:energy_diff}
\end{figure}

Many of the comparisons and insights we present in this section stem from the total energy
results of our CPS ansatz, which are displayed in Table \ref{tab:energies} and Figure \ref{fig:total_energy}.
As we described above, working with the CPS wavefunction in the variational Monte Carlo framework, it is simple to constrain the value
of the total system's $S_z$ spin, and so we are able to probe the lowest energy state in each $S_z$ sector.
As seen in Figure \ref{fig:total_energy}, the minimum energy as a function of $S_z$ is nearly parabolic, as found in previous DMRG
calculations \cite{SCHNACK:2003:mo72fe30_dmrg}, and thus the CPS energies provide another wavefunction-based verification of
the qualitative correctness of the rotational band model (RBM).
The agreement is not quantitative, however, and Figure \ref{fig:energy_diff} shows the deviation of the raw CPS energies when
we try to fit them to the RBM form in Eq.\ (\ref{eqn:rbm_energy}).
We see that except for the region near 1/3 of the maximum magnetization, the differences between the CPS and RBM energies can
be fit closely by a cubic correction, which is not surprising as cubic terms are the leading order terms neglected by the RBM.
The sharp change in the deviations near 1/3 of saturation is more interesting, however, as it is responsible for creating
the $B_{sat}/3$ anomalies that cannot be predicted by the RBM.
We will discuss these anomalies and the origins of the energy deviations responsible in Sections \ref{sec:properties} and
\ref{sec:phase_transition}.

\begin{table*}
\centering
\begin{minipage}{0.85\textwidth}
\caption{
         Ground state energies of CPS and the RBM, in units of $J$, for the $S=5/2$ Heisenberg model on the
         icosidodecahedron for different total $S_z$ sectors of Hilbert space.
         The raw energies of the CPS wavefunction are given, as well as the
         energies produced by fits to the RBM using CPS energies, DMRG energies
         \cite{SCHNACK:2003:mo72fe30_dmrg}, and experimental magnetizations \cite{SCHNACK:2001:rot_band_model}.
         See Section \ref{sec:total_energies}.
        }
\label{tab:energies}
\begin{tabular}{  r  r@{.}l   r@{.}l   r@{.}l   r@{.}l  r  r@{.}l   r@{.}l   r@{.}l   r@{.}l  }
\hline\hline
\hspace{0mm} $S_z$ \hspace{0mm} &
\multicolumn{2}{ c }{ \hspace{0mm} CPS    \hspace{0mm} } &
\multicolumn{2}{ c }{ \hspace{0mm} Fit    \hspace{0mm} } &
\multicolumn{2}{ c }{ \hspace{0mm} DMRG   \hspace{0mm} } &
\multicolumn{2}{ c }{ \hspace{0mm} Exp.   \hspace{0mm} } &
\hspace{0mm} $S_z$ \hspace{0mm} &
\multicolumn{2}{ c }{ \hspace{0mm} CPS    \hspace{0mm} } &
\multicolumn{2}{ c }{ \hspace{0mm} Fit    \hspace{0mm} } &
\multicolumn{2}{ c }{ \hspace{0mm} DMRG   \hspace{0mm} } &
\multicolumn{2}{ c }{ \hspace{0mm} Exp.   \hspace{0mm} } \\
\hline
 \hspace{0mm}  0   \hspace{0mm} & \hspace{1mm}    -216&25 \hspace{1mm} & \hspace{1mm}  -216&25  \hspace{1mm} & \hspace{1mm}  -210&55 \hspace{1mm} & \hspace{1mm}   -216&65  \hspace{1mm} & \hspace{0mm} 38   \hspace{0mm} & \hspace{1mm}     -61&23 \hspace{1mm} & \hspace{1mm}   -62&52  \hspace{1mm} & \hspace{1mm}   -58&15 \hspace{1mm} & \hspace{1mm}    -62&77  \hspace{1mm} \\
 \hspace{0mm}  1   \hspace{0mm} & \hspace{1mm}    -216&14 \hspace{1mm} & \hspace{1mm}  -216&04  \hspace{1mm} & \hspace{1mm}  -210&35 \hspace{1mm} & \hspace{1mm}   -216&44  \hspace{1mm} & \hspace{0mm} 39   \hspace{0mm} & \hspace{1mm}     -52&91 \hspace{1mm} & \hspace{1mm}   -54&43  \hspace{1mm} & \hspace{1mm}   -50&13 \hspace{1mm} & \hspace{1mm}    -54&67  \hspace{1mm} \\
 \hspace{0mm}  2   \hspace{0mm} & \hspace{1mm}    -215&80 \hspace{1mm} & \hspace{1mm}  -215&63  \hspace{1mm} & \hspace{1mm}  -209&93 \hspace{1mm} & \hspace{1mm}   -216&03  \hspace{1mm} & \hspace{0mm} 40   \hspace{0mm} & \hspace{1mm}     -44&39 \hspace{1mm} & \hspace{1mm}   -46&14  \hspace{1mm} & \hspace{1mm}   -41&90 \hspace{1mm} & \hspace{1mm}    -46&36  \hspace{1mm} \\
 \hspace{0mm}  3   \hspace{0mm} & \hspace{1mm}    -215&23 \hspace{1mm} & \hspace{1mm}  -215&00  \hspace{1mm} & \hspace{1mm}  -209&32 \hspace{1mm} & \hspace{1mm}   -215&40  \hspace{1mm} & \hspace{0mm} 41   \hspace{0mm} & \hspace{1mm}     -35&68 \hspace{1mm} & \hspace{1mm}   -37&63  \hspace{1mm} & \hspace{1mm}   -33&47 \hspace{1mm} & \hspace{1mm}    -37&85  \hspace{1mm} \\
 \hspace{0mm}  4   \hspace{0mm} & \hspace{1mm}    -214&45 \hspace{1mm} & \hspace{1mm}  -214&18  \hspace{1mm} & \hspace{1mm}  -208&49 \hspace{1mm} & \hspace{1mm}   -214&57  \hspace{1mm} & \hspace{0mm} 42   \hspace{0mm} & \hspace{1mm}     -26&78 \hspace{1mm} & \hspace{1mm}   -28&92  \hspace{1mm} & \hspace{1mm}   -24&83 \hspace{1mm} & \hspace{1mm}    -29&13  \hspace{1mm} \\
 \hspace{0mm}  5   \hspace{0mm} & \hspace{1mm}    -213&43 \hspace{1mm} & \hspace{1mm}  -213&14  \hspace{1mm} & \hspace{1mm}  -207&47 \hspace{1mm} & \hspace{1mm}   -213&53  \hspace{1mm} & \hspace{0mm} 43   \hspace{0mm} & \hspace{1mm}     -17&67 \hspace{1mm} & \hspace{1mm}   -20&00  \hspace{1mm} & \hspace{1mm}   -15&99 \hspace{1mm} & \hspace{1mm}    -20&20  \hspace{1mm} \\
 \hspace{0mm}  6   \hspace{0mm} & \hspace{1mm}    -212&20 \hspace{1mm} & \hspace{1mm}  -211&89  \hspace{1mm} & \hspace{1mm}  -206&23 \hspace{1mm} & \hspace{1mm}   -212&29  \hspace{1mm} & \hspace{0mm} 44   \hspace{0mm} & \hspace{1mm}      -8&37 \hspace{1mm} & \hspace{1mm}   -10&87  \hspace{1mm} & \hspace{1mm}    -6&94 \hspace{1mm} & \hspace{1mm}    -11&06  \hspace{1mm} \\
 \hspace{0mm}  7   \hspace{0mm} & \hspace{1mm}    -210&74 \hspace{1mm} & \hspace{1mm}  -210&44  \hspace{1mm} & \hspace{1mm}  -204&79 \hspace{1mm} & \hspace{1mm}   -210&83  \hspace{1mm} & \hspace{0mm} 45   \hspace{0mm} & \hspace{1mm}       1&12 \hspace{1mm} & \hspace{1mm}    -1&53  \hspace{1mm} & \hspace{1mm}     2&31 \hspace{1mm} & \hspace{1mm}     -1&71  \hspace{1mm} \\
 \hspace{0mm}  8   \hspace{0mm} & \hspace{1mm}    -209&06 \hspace{1mm} & \hspace{1mm}  -208&78  \hspace{1mm} & \hspace{1mm}  -203&15 \hspace{1mm} & \hspace{1mm}   -209&17  \hspace{1mm} & \hspace{0mm} 46   \hspace{0mm} & \hspace{1mm}      10&81 \hspace{1mm} & \hspace{1mm}     8&01  \hspace{1mm} & \hspace{1mm}    11&77 \hspace{1mm} & \hspace{1mm}      7&84  \hspace{1mm} \\
 \hspace{0mm}  9   \hspace{0mm} & \hspace{1mm}    -207&16 \hspace{1mm} & \hspace{1mm}  -206&91  \hspace{1mm} & \hspace{1mm}  -201&30 \hspace{1mm} & \hspace{1mm}   -207&30  \hspace{1mm} & \hspace{0mm} 47   \hspace{0mm} & \hspace{1mm}      20&70 \hspace{1mm} & \hspace{1mm}    17&76  \hspace{1mm} & \hspace{1mm}    21&44 \hspace{1mm} & \hspace{1mm}     17&60  \hspace{1mm} \\
 \hspace{0mm} 10   \hspace{0mm} & \hspace{1mm}    -205&05 \hspace{1mm} & \hspace{1mm}  -204&84  \hspace{1mm} & \hspace{1mm}  -199&24 \hspace{1mm} & \hspace{1mm}   -205&23  \hspace{1mm} & \hspace{0mm} 48   \hspace{0mm} & \hspace{1mm}      30&77 \hspace{1mm} & \hspace{1mm}    27&72  \hspace{1mm} & \hspace{1mm}    31&31 \hspace{1mm} & \hspace{1mm}     27&57  \hspace{1mm} \\
 \hspace{0mm} 11   \hspace{0mm} & \hspace{1mm}    -202&71 \hspace{1mm} & \hspace{1mm}  -202&56  \hspace{1mm} & \hspace{1mm}  -196&98 \hspace{1mm} & \hspace{1mm}   -202&94  \hspace{1mm} & \hspace{0mm} 49   \hspace{0mm} & \hspace{1mm}      41&05 \hspace{1mm} & \hspace{1mm}    37&88  \hspace{1mm} & \hspace{1mm}    41&39 \hspace{1mm} & \hspace{1mm}     37&74  \hspace{1mm} \\
 \hspace{0mm} 12   \hspace{0mm} & \hspace{1mm}    -200&16 \hspace{1mm} & \hspace{1mm}  -200&07  \hspace{1mm} & \hspace{1mm}  -194&51 \hspace{1mm} & \hspace{1mm}   -200&45  \hspace{1mm} & \hspace{0mm} 50   \hspace{0mm} & \hspace{1mm}      51&52 \hspace{1mm} & \hspace{1mm}    48&26  \hspace{1mm} & \hspace{1mm}    51&67 \hspace{1mm} & \hspace{1mm}     48&13  \hspace{1mm} \\
 \hspace{0mm} 13   \hspace{0mm} & \hspace{1mm}    -197&38 \hspace{1mm} & \hspace{1mm}  -197&37  \hspace{1mm} & \hspace{1mm}  -191&84 \hspace{1mm} & \hspace{1mm}   -197&75  \hspace{1mm} & \hspace{0mm} 51   \hspace{0mm} & \hspace{1mm}      62&18 \hspace{1mm} & \hspace{1mm}    58&84  \hspace{1mm} & \hspace{1mm}    62&16 \hspace{1mm} & \hspace{1mm}     58&72  \hspace{1mm} \\
 \hspace{0mm} 14   \hspace{0mm} & \hspace{1mm}    -194&39 \hspace{1mm} & \hspace{1mm}  -194&47  \hspace{1mm} & \hspace{1mm}  -188&96 \hspace{1mm} & \hspace{1mm}   -194&84  \hspace{1mm} & \hspace{0mm} 52   \hspace{0mm} & \hspace{1mm}      73&04 \hspace{1mm} & \hspace{1mm}    69&62  \hspace{1mm} & \hspace{1mm}    72&86 \hspace{1mm} & \hspace{1mm}     69&52  \hspace{1mm} \\
 \hspace{0mm} 15   \hspace{0mm} & \hspace{1mm}    -191&17 \hspace{1mm} & \hspace{1mm}  -191&35  \hspace{1mm} & \hspace{1mm}  -185&87 \hspace{1mm} & \hspace{1mm}   -191&73  \hspace{1mm} & \hspace{0mm} 53   \hspace{0mm} & \hspace{1mm}      84&09 \hspace{1mm} & \hspace{1mm}    80&62  \hspace{1mm} & \hspace{1mm}    83&76 \hspace{1mm} & \hspace{1mm}     80&52  \hspace{1mm} \\
 \hspace{0mm} 16   \hspace{0mm} & \hspace{1mm}    -187&76 \hspace{1mm} & \hspace{1mm}  -188&04  \hspace{1mm} & \hspace{1mm}  -182&58 \hspace{1mm} & \hspace{1mm}   -188&41  \hspace{1mm} & \hspace{0mm} 54   \hspace{0mm} & \hspace{1mm}      95&34 \hspace{1mm} & \hspace{1mm}    91&82  \hspace{1mm} & \hspace{1mm}    94&86 \hspace{1mm} & \hspace{1mm}     91&74  \hspace{1mm} \\
 \hspace{0mm} 17   \hspace{0mm} & \hspace{1mm}    -184&11 \hspace{1mm} & \hspace{1mm}  -184&51  \hspace{1mm} & \hspace{1mm}  -179&08 \hspace{1mm} & \hspace{1mm}   -184&88  \hspace{1mm} & \hspace{0mm} 55   \hspace{0mm} & \hspace{1mm}     106&78 \hspace{1mm} & \hspace{1mm}   103&23  \hspace{1mm} & \hspace{1mm}   106&18 \hspace{1mm} & \hspace{1mm}    103&16  \hspace{1mm} \\
 \hspace{0mm} 18   \hspace{0mm} & \hspace{1mm}    -180&24 \hspace{1mm} & \hspace{1mm}  -180&77  \hspace{1mm} & \hspace{1mm}  -175&38 \hspace{1mm} & \hspace{1mm}   -181&14  \hspace{1mm} & \hspace{0mm} 56   \hspace{0mm} & \hspace{1mm}     118&41 \hspace{1mm} & \hspace{1mm}   114&85  \hspace{1mm} & \hspace{1mm}   117&69 \hspace{1mm} & \hspace{1mm}    114&79  \hspace{1mm} \\
 \hspace{0mm} 19   \hspace{0mm} & \hspace{1mm}    -176&17 \hspace{1mm} & \hspace{1mm}  -176&83  \hspace{1mm} & \hspace{1mm}  -171&47 \hspace{1mm} & \hspace{1mm}   -177&19  \hspace{1mm} & \hspace{0mm} 57   \hspace{0mm} & \hspace{1mm}     130&24 \hspace{1mm} & \hspace{1mm}   126&68  \hspace{1mm} & \hspace{1mm}   129&42 \hspace{1mm} & \hspace{1mm}    126&62  \hspace{1mm} \\
 \hspace{0mm} 20   \hspace{0mm} & \hspace{1mm}    -172&00 \hspace{1mm} & \hspace{1mm}  -172&68  \hspace{1mm} & \hspace{1mm}  -167&36 \hspace{1mm} & \hspace{1mm}   -173&04  \hspace{1mm} & \hspace{0mm} 58   \hspace{0mm} & \hspace{1mm}     142&26 \hspace{1mm} & \hspace{1mm}   138&71  \hspace{1mm} & \hspace{1mm}   141&34 \hspace{1mm} & \hspace{1mm}    138&67  \hspace{1mm} \\
 \hspace{0mm} 21   \hspace{0mm} & \hspace{1mm}    -168&22 \hspace{1mm} & \hspace{1mm}  -168&33  \hspace{1mm} & \hspace{1mm}  -163&04 \hspace{1mm} & \hspace{1mm}   -168&68  \hspace{1mm} & \hspace{0mm} 59   \hspace{0mm} & \hspace{1mm}     154&47 \hspace{1mm} & \hspace{1mm}   150&95  \hspace{1mm} & \hspace{1mm}   153&48 \hspace{1mm} & \hspace{1mm}    150&92  \hspace{1mm} \\
 \hspace{0mm} 22   \hspace{0mm} & \hspace{1mm}    -163&79 \hspace{1mm} & \hspace{1mm}  -163&76  \hspace{1mm} & \hspace{1mm}  -158&52 \hspace{1mm} & \hspace{1mm}   -164&11  \hspace{1mm} & \hspace{0mm} 60   \hspace{0mm} & \hspace{1mm}     166&88 \hspace{1mm} & \hspace{1mm}   163&39  \hspace{1mm} & \hspace{1mm}   165&82 \hspace{1mm} & \hspace{1mm}    163&38  \hspace{1mm} \\
 \hspace{0mm} 23   \hspace{0mm} & \hspace{1mm}    -159&20 \hspace{1mm} & \hspace{1mm}  -158&99  \hspace{1mm} & \hspace{1mm}  -153&79 \hspace{1mm} & \hspace{1mm}   -159&33  \hspace{1mm} & \hspace{0mm} 61   \hspace{0mm} & \hspace{1mm}     179&47 \hspace{1mm} & \hspace{1mm}   176&05  \hspace{1mm} & \hspace{1mm}   178&36 \hspace{1mm} & \hspace{1mm}    176&05  \hspace{1mm} \\
 \hspace{0mm} 24   \hspace{0mm} & \hspace{1mm}    -154&42 \hspace{1mm} & \hspace{1mm}  -154&01  \hspace{1mm} & \hspace{1mm}  -148&85 \hspace{1mm} & \hspace{1mm}   -154&35  \hspace{1mm} & \hspace{0mm} 62   \hspace{0mm} & \hspace{1mm}     192&26 \hspace{1mm} & \hspace{1mm}   188&91  \hspace{1mm} & \hspace{1mm}   191&12 \hspace{1mm} & \hspace{1mm}    188&92  \hspace{1mm} \\
 \hspace{0mm} 25   \hspace{0mm} & \hspace{1mm}    -149&76 \hspace{1mm} & \hspace{1mm}  -148&83  \hspace{1mm} & \hspace{1mm}  -143&71 \hspace{1mm} & \hspace{1mm}   -149&16  \hspace{1mm} & \hspace{0mm} 63   \hspace{0mm} & \hspace{1mm}     205&23 \hspace{1mm} & \hspace{1mm}   201&98  \hspace{1mm} & \hspace{1mm}   204&07 \hspace{1mm} & \hspace{1mm}    202&01  \hspace{1mm} \\
 \hspace{0mm} 26   \hspace{0mm} & \hspace{1mm}    -144&40 \hspace{1mm} & \hspace{1mm}  -143&43  \hspace{1mm} & \hspace{1mm}  -138&36 \hspace{1mm} & \hspace{1mm}   -143&76  \hspace{1mm} & \hspace{0mm} 64   \hspace{0mm} & \hspace{1mm}     218&40 \hspace{1mm} & \hspace{1mm}   215&26  \hspace{1mm} & \hspace{1mm}   217&24 \hspace{1mm} & \hspace{1mm}    215&30  \hspace{1mm} \\
 \hspace{0mm} 27   \hspace{0mm} & \hspace{1mm}    -138&87 \hspace{1mm} & \hspace{1mm}  -137&83  \hspace{1mm} & \hspace{1mm}  -132&81 \hspace{1mm} & \hspace{1mm}   -138&15  \hspace{1mm} & \hspace{0mm} 65   \hspace{0mm} & \hspace{1mm}     231&75 \hspace{1mm} & \hspace{1mm}   228&74  \hspace{1mm} & \hspace{1mm}   230&60 \hspace{1mm} & \hspace{1mm}    228&80  \hspace{1mm} \\
 \hspace{0mm} 28   \hspace{0mm} & \hspace{1mm}    -133&01 \hspace{1mm} & \hspace{1mm}  -132&02  \hspace{1mm} & \hspace{1mm}  -127&05 \hspace{1mm} & \hspace{1mm}   -132&34  \hspace{1mm} & \hspace{0mm} 66   \hspace{0mm} & \hspace{1mm}     245&30 \hspace{1mm} & \hspace{1mm}   242&44  \hspace{1mm} & \hspace{1mm}   244&18 \hspace{1mm} & \hspace{1mm}    242&50  \hspace{1mm} \\
 \hspace{0mm} 29   \hspace{0mm} & \hspace{1mm}    -126&86 \hspace{1mm} & \hspace{1mm}  -126&01  \hspace{1mm} & \hspace{1mm}  -121&09 \hspace{1mm} & \hspace{1mm}   -126&31  \hspace{1mm} & \hspace{0mm} 67   \hspace{0mm} & \hspace{1mm}     259&02 \hspace{1mm} & \hspace{1mm}   256&33  \hspace{1mm} & \hspace{1mm}   257&96 \hspace{1mm} & \hspace{1mm}    256&42  \hspace{1mm} \\
 \hspace{0mm} 30   \hspace{0mm} & \hspace{1mm}    -120&43 \hspace{1mm} & \hspace{1mm}  -119&78  \hspace{1mm} & \hspace{1mm}  -114&92 \hspace{1mm} & \hspace{1mm}   -120&08  \hspace{1mm} & \hspace{0mm} 68   \hspace{0mm} & \hspace{1mm}     272&93 \hspace{1mm} & \hspace{1mm}   270&44  \hspace{1mm} & \hspace{1mm}   271&94 \hspace{1mm} & \hspace{1mm}    270&54  \hspace{1mm} \\
 \hspace{0mm} 31   \hspace{0mm} & \hspace{1mm}    -113&78 \hspace{1mm} & \hspace{1mm}  -113&35  \hspace{1mm} & \hspace{1mm}  -108&54 \hspace{1mm} & \hspace{1mm}   -113&65  \hspace{1mm} & \hspace{0mm} 69   \hspace{0mm} & \hspace{1mm}     287&03 \hspace{1mm} & \hspace{1mm}   284&76  \hspace{1mm} & \hspace{1mm}   286&13 \hspace{1mm} & \hspace{1mm}    284&87  \hspace{1mm} \\
 \hspace{0mm} 32   \hspace{0mm} & \hspace{1mm}    -106&92 \hspace{1mm} & \hspace{1mm}  -106&71  \hspace{1mm} & \hspace{1mm}  -101&96 \hspace{1mm} & \hspace{1mm}   -107&00  \hspace{1mm} & \hspace{0mm} 70   \hspace{0mm} & \hspace{1mm}     301&30 \hspace{1mm} & \hspace{1mm}   299&28  \hspace{1mm} & \hspace{1mm}   300&53 \hspace{1mm} & \hspace{1mm}    299&40  \hspace{1mm} \\
 \hspace{0mm} 33   \hspace{0mm} & \hspace{1mm}     -99&82 \hspace{1mm} & \hspace{1mm}   -99&87  \hspace{1mm} & \hspace{1mm}   -95&17 \hspace{1mm} & \hspace{1mm}   -100&15  \hspace{1mm} & \hspace{0mm} 71   \hspace{0mm} & \hspace{1mm}     315&75 \hspace{1mm} & \hspace{1mm}   314&01  \hspace{1mm} & \hspace{1mm}   315&13 \hspace{1mm} & \hspace{1mm}    314&15  \hspace{1mm} \\
 \hspace{0mm} 34   \hspace{0mm} & \hspace{1mm}     -92&50 \hspace{1mm} & \hspace{1mm}   -92&81  \hspace{1mm} & \hspace{1mm}   -88&18 \hspace{1mm} & \hspace{1mm}    -93&09  \hspace{1mm} & \hspace{0mm} 72   \hspace{0mm} & \hspace{1mm}     330&35 \hspace{1mm} & \hspace{1mm}   328&94  \hspace{1mm} & \hspace{1mm}   329&94 \hspace{1mm} & \hspace{1mm}    329&10  \hspace{1mm} \\
 \hspace{0mm} 35   \hspace{0mm} & \hspace{1mm}     -84&98 \hspace{1mm} & \hspace{1mm}   -85&55  \hspace{1mm} & \hspace{1mm}   -80&98 \hspace{1mm} & \hspace{1mm}    -85&82  \hspace{1mm} & \hspace{0mm} 73   \hspace{0mm} & \hspace{1mm}     345&18 \hspace{1mm} & \hspace{1mm}   344&09  \hspace{1mm} & \hspace{1mm}   344&95 \hspace{1mm} & \hspace{1mm}    344&26  \hspace{1mm} \\
 \hspace{0mm} 36   \hspace{0mm} & \hspace{1mm}     -77&26 \hspace{1mm} & \hspace{1mm}   -78&08  \hspace{1mm} & \hspace{1mm}   -73&58 \hspace{1mm} & \hspace{1mm}    -78&34  \hspace{1mm} & \hspace{0mm} 74   \hspace{0mm} & \hspace{1mm}     360&00 \hspace{1mm} & \hspace{1mm}   359&44  \hspace{1mm} & \hspace{1mm}   360&17 \hspace{1mm} & \hspace{1mm}    359&63  \hspace{1mm} \\
 \hspace{0mm} 37   \hspace{0mm} & \hspace{1mm}     -69&33 \hspace{1mm} & \hspace{1mm}   -70&41  \hspace{1mm} & \hspace{1mm}   -65&97 \hspace{1mm} & \hspace{1mm}    -70&66  \hspace{1mm} & \hspace{0mm} 75   \hspace{0mm} & \hspace{1mm}     375&00 \hspace{1mm} & \hspace{1mm}   375&00  \hspace{1mm} & \hspace{1mm}   375&60 \hspace{1mm} & \hspace{1mm}    375&20  \hspace{1mm} \\
\hline\hline
\end{tabular}
\end{minipage}
\end{table*}

\clearpage

In the inset to Figure \ref{fig:total_energy}, we see that our CPS calculations produce 
superior variational energies as compared to DMRG.
In addition to producing superior variational energies, a fit of the CPS energies to the lowest band of the RBM
produces band parameters ($D=6.22, \gamma=1.07$) that resemble much more closely the band parameters fitted to experimental magnetization
data ($D=6.23, \gamma=1.07$) \cite{SCHNACK:2001:rot_band_model} than those produced by a fit to the DMRG energies
($D=6.17, \gamma=1.05$) \cite{SCHNACK:2003:mo72fe30_dmrg}.

In addition to RBM comparisons, we may compare the CPS singlet-triplet gap with the 
singlet-triplet gap  measured by neutron scattering. 
Using the value of $J\approx0.134$ meV \cite{MULLER:2001:giant_keplerate,SCHNACK:2001:rot_band_model}, we find that our CPS
calculations predict a gap of 0.015 meV, which is smaller than the RBM result of 0.027 meV and significantly smaller than the
0.091 meV gap derived from neutron scattering data \cite{FU:2010:neutron_scatter}.
Since both the $S_z=0$ and $S_z=1$ energies are upper bounds, this suggests that the CPS $S_z=0$ ground-state energy, although
an improvement over the DMRG energy,  must still be relatively too high. This motivates further improvements in the CPS ansatz, 
possibilities for which we mention in the conclusion.

\begin{figure}[t]
\centering
\includegraphics[width=8.5cm,angle=0]{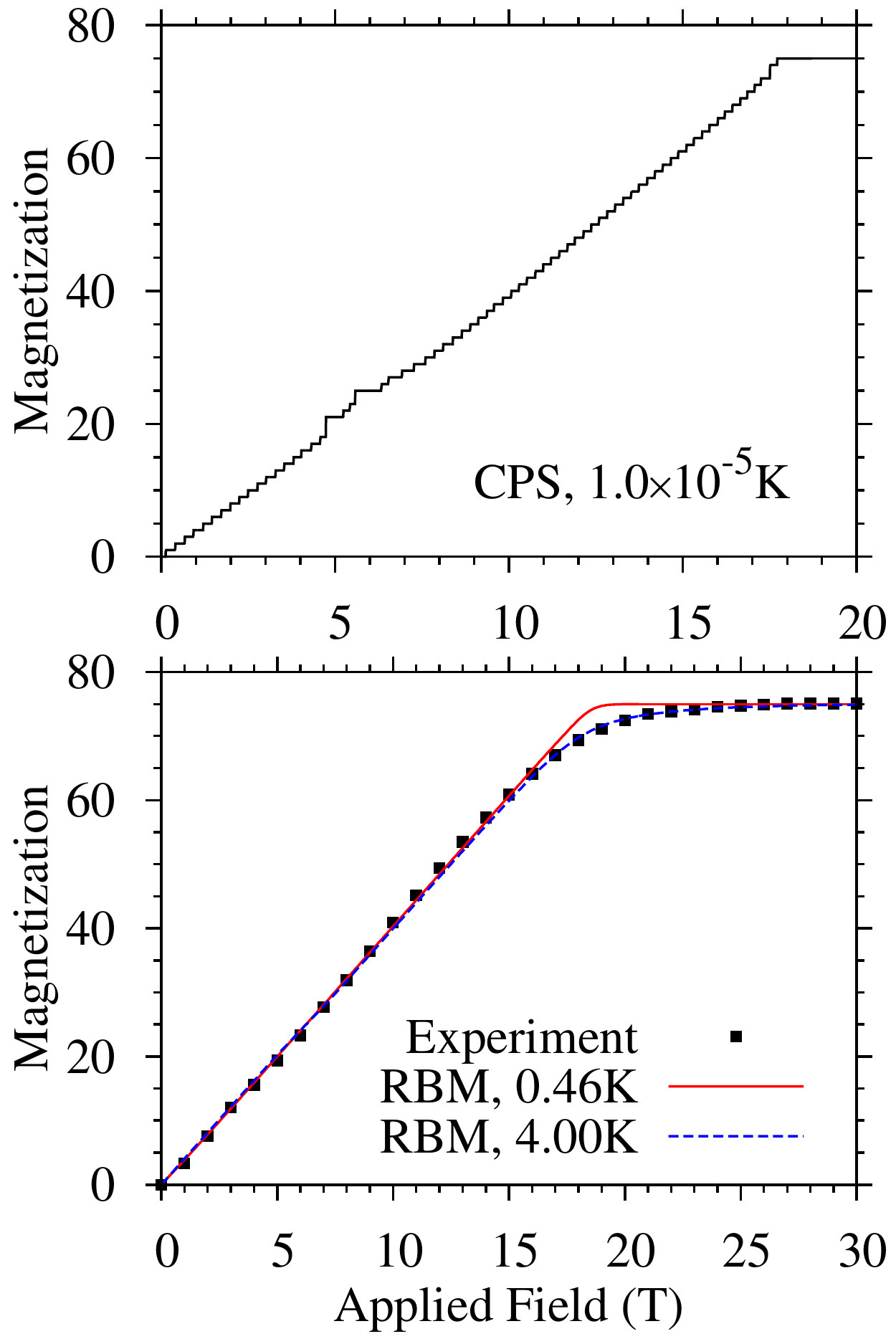}
\caption{Total $z$ magnetizations at different applied field strengths of the $S=5/2$ icosidodecahedron.
         In the top panel we plot the low temperature limit of the magnetization curve derived from the
         CPS wavefunction energies.
         In the lower panel we plot finite temperature curves for the rotational band model using
         an ab initio parameterization based on the fit to the CPS energies (see Figure
         \ref{fig:total_energy}),
         as well as the experimental results of Schnack et al \cite{SCHNACK:2001:rot_band_model}.
         See Section \ref{sec:properties}.
        }
\label{fig:magnetization}
\end{figure}

\subsection{Properties}
\label{sec:properties}

Using the raw CPS energies or their fit to the rotational band model, we can evaluate the magnetization $M$,
differential magnetization $dM/dB$, and heat capacity $C_p$ of the $S=5/2$ icosidodecahedron, as functions
of the applied field $B$.
Note that in our calculations of $dM/dB$ and $C_p$ we do not include the effects of excited states other
than the lowest state in each spin-sector.
Indeed, we show that including only the lowest spin-state contributions already produces much of the
anomalous behavior common to spin systems built of corner sharing triangles.
Furthermore, in the case of $dM/dB$, the neglect of low-lying singlets is probably a reasonable
approximation at low temperatures, as such states do not contribute directly.

To make quantitative predictions, we have taken \cite{MULLER:2001:giant_keplerate,SCHNACK:2001:rot_band_model}
the interaction strength as $J/k_B=1.566K$ and the spectroscopic splitting factor as $g=1.974$.
We begin by considering the magnetization curve at finite temperatures, for which experimental results can be
matched closely by the RBM \cite{SCHNACK:2001:rot_band_model}.
As may be expected by the similarity between the experimental and ab initio CPS fittings of the band model parameters
discussed previously, the magnetization curve obtained from the
CPS-parameterized RBM also 
matches the experimental magnetization curve closely,
as seen in the bottom panel of Figure \ref{fig:magnetization}.
More interesting, however,  is the zero temperature limit of the magnetization curve, shown in the upper panel of Figure \ref{fig:magnetization},
where we see that the icosidodecahedron has anomalies in its magnetization staircase at field strengths near 1/3 of the
saturation field strength $B_{sat}=17.7T$.
The staircase anomaly directly reflects the deviations in the ground-state energy levels, for each $S_z$, from the parabolic trend assumed by the RBM.
We leave the discussion of the physical interpretation of the energy level deviations and corresponding
staircase anomaly to section  \ref{sec:phase_transition}. Here we will show that this feature of the
ground-state spectrum is sufficient to reproduce most of the unusual properties of the magnet near $B_{sat}/3$, without
the need to explicitly consider other excited states.

\begin{figure}[t!]
\includegraphics[width=8.0cm,angle=0]{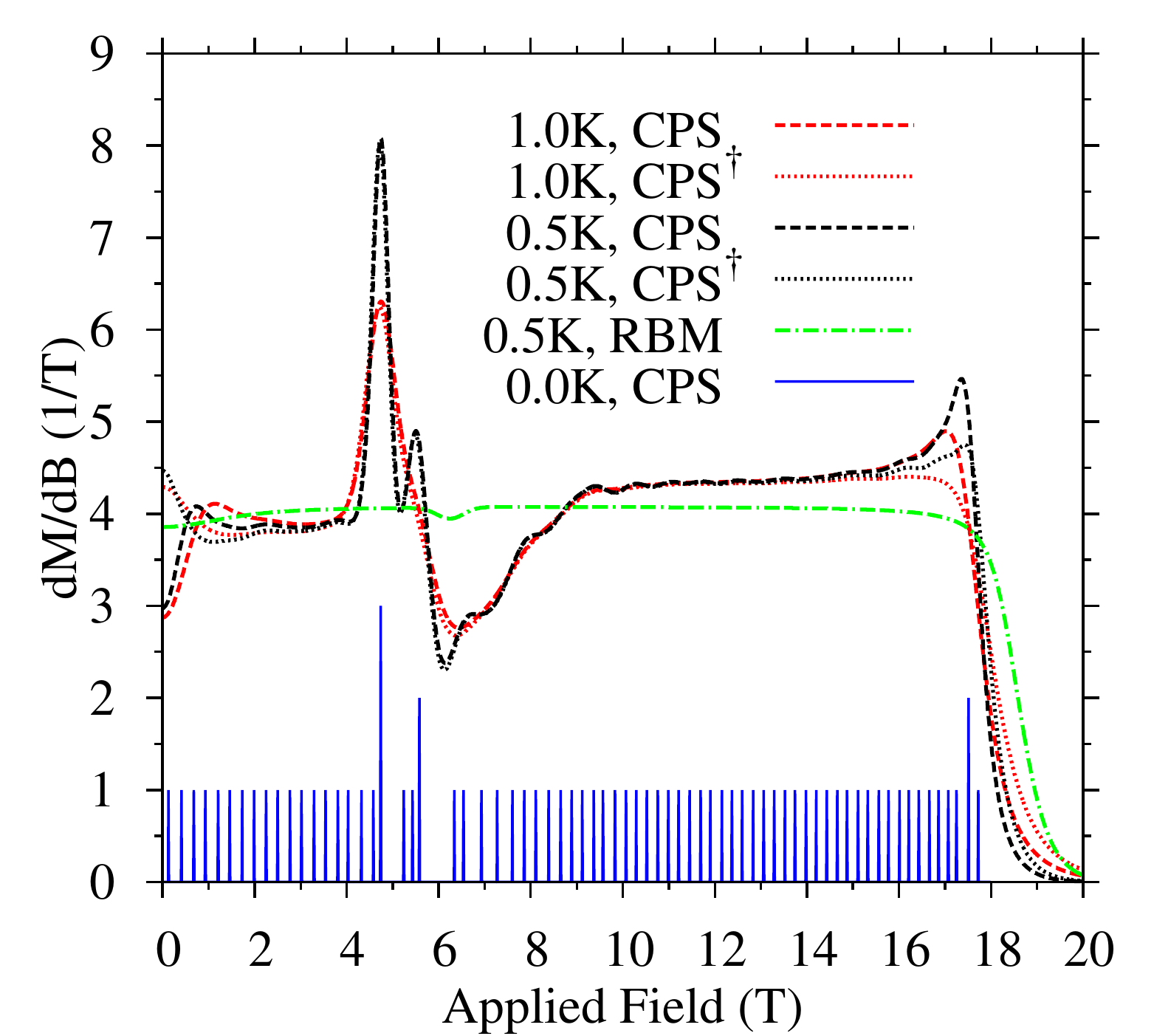}
\caption{Differential susceptibility $dM/dB$ as a function of the applied field strength in Teslas.
         Results for the CPS wavefunction are shown both for the case when the ground state in each $S_z$ sector
         is assumed to be non-degenerate (CPS) and when each $S_z$ ground state is assumed to have the same
         degeneracy as the corresponding state in the RBM ($\mathrm{CPS^\dagger}$).
         For comparison, we also show the susceptibility derived from the first rotational band of the RBM with
         its experimentally derived parameterization \cite{SCHNACK:2001:rot_band_model}.
         Note that in the zero temperature case the delta functions that make up the $dM/dB$ curve have been
         scaled arbitrarily to show the number of magnetization levels ascended at each ``step'', so for
         example the line just below 5$T$ represents a magnetization change of 3, while most lines represent
         a change of 1.
         See Section \ref{sec:properties}.
        }
\label{fig:dmdb}
\includegraphics[width=8.0cm,angle=0]{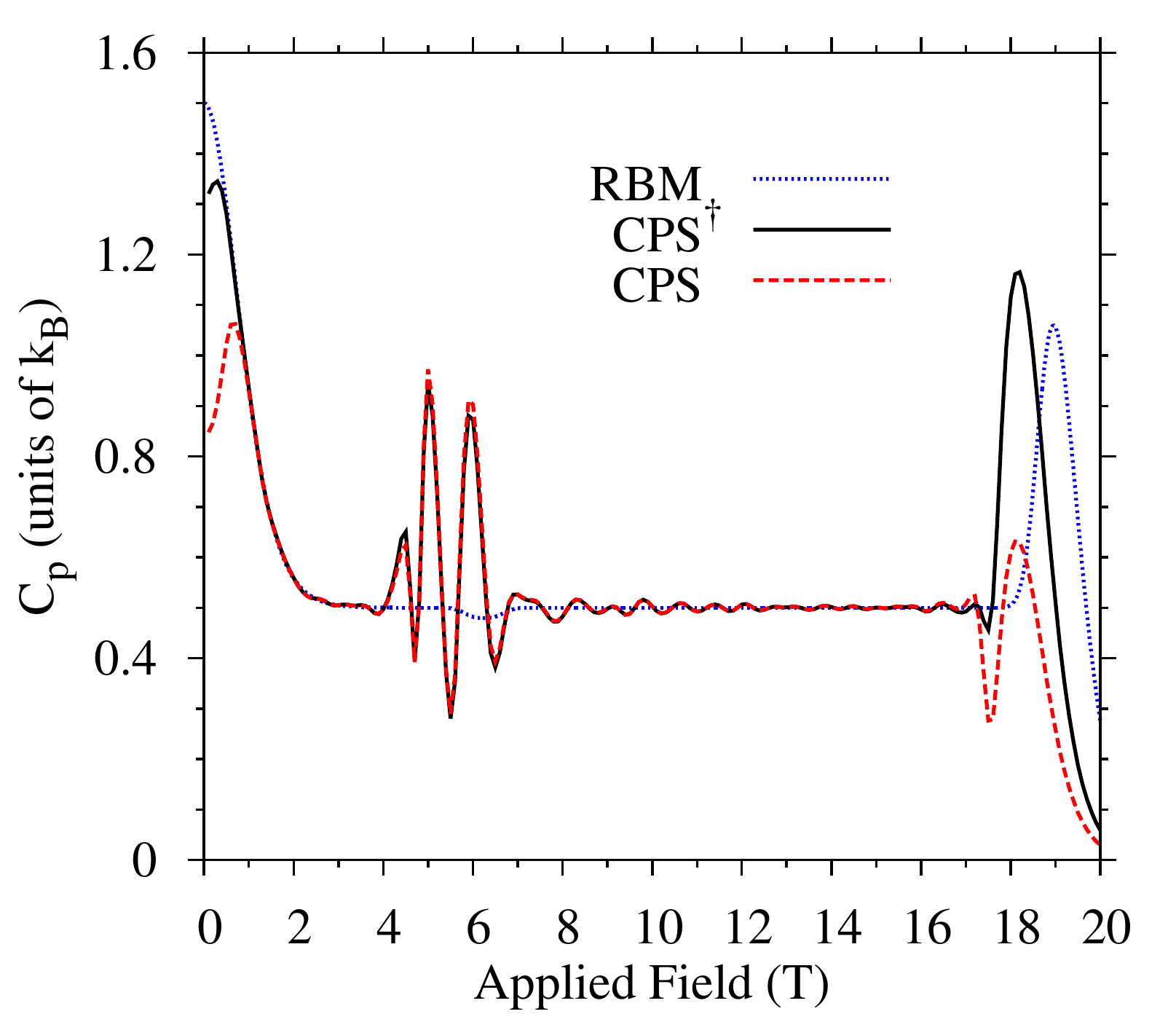}
\caption{Heat capacity as a function of the applied field strength in Teslas at $T$=0.4K.
         Results for the CPS wavefunction are shown both for the case when the ground state in each $S_z$ sector
         is assumed to be non-degenerate (CPS) and when each $S_z$ ground state is assumed to have the same
         degeneracy as the corresponding state in the RBM ($\mathrm{CPS^\dagger}$).
         For comparison, we also show the heat capacity derived from the first rotational band of the RBM with
         its experimentally derived parameterization \cite{SCHNACK:2001:rot_band_model}.
         See Section \ref{sec:properties}.
        }
\label{fig:heat_capacity}
\end{figure}

The differential susceptibility derived from the CPS energies is shown in Figure \ref{fig:dmdb}.
We see that there is a sharp rise followed by a depression in the differential
susceptibility, which in the case of the 0.5K results can be clearly associated
with the staircase anomalies, which show up as gaps in the delta function progression of the 0K $dM/dB$ curve.
Note that the area in the trough is greater than that in the peak, which, in conjunction with inhomogeneities in
the interactions \cite{SCHRODER:2008:random_exchange} of the $\mathrm{Mo_{72}Fe_{30}}$ compound that could smear the
features together, may explain why only a broad trough is seen in experimental measurements
\cite{SCHRODER:2005:diff_susc,SCHRODER:2008:random_exchange}.
In contrast to the CPS results, the RBM predicts only a very small dip in the $dM/dB$ curve near $B_{sat}/3$.

As for the case of differential susceptibility, the heat capacity also shows a distinct feature near $B_{sat}/3$,
even when the low-lying excited states are ignored as in our CPS calculations.
(Note that to the best of our knowledge, detailed measurements of the heat capacity are not yet available).
As shown in Figure \ref{fig:heat_capacity}, the heat capacity derived from the CPS energies oscillates near the
staircase anomaly, whereas the heat capacity derived from the RBM shows only a small dip in this region.
Note that the oscillations are present and essentially the same both when the CPS ground states are assumed to be
non-degenerate and when they are assumed to have the same degeneracies as the corresponding states in the RBM.
This  offers  reason to expect that the feature would be robust to the inclusion of additional excited states.


\begin{table}[h!]
\centering
\caption{
         Averages of the dot product $\vec{S}_i \cdot \vec{S}_j / (|\vec{S}_i| |\vec{S}_j|)$
         for different choices of the sublattices (A,B,C) for sites $i$ and $j$.
         The CPS results are for the wavefunction with zero total $S_z$, while the numbers
         for the classical Heisenberg model correspond to zero applied field.
         The abbreviation n.n. stands for nearest neighbor.
         See Section \ref{sec:spin_correlations}.
        }
\label{tab:spin_corr}
\begin{tabular}{  c c c   r@{.}l   r@{.}l  }
\hline\hline
\hspace{0mm} average type \hspace{0mm} &
\hspace{2mm} $i$ \hspace{2mm} &
\hspace{2mm} $j$ \hspace{2mm} &
\multicolumn{2}{ c }{ \hspace{0mm} CPS        \hspace{0mm} } &
\multicolumn{2}{ c }{ \hspace{0mm} Classical  \hspace{0mm} } \\
\hline
 \hspace{2mm} all       \hspace{2mm}  & A & A & \hspace{3mm}  0&79  \hspace{3mm} & \hspace{2mm}  1&00 \hspace{2mm} \\
 \hspace{2mm} all       \hspace{2mm}  & B & B & \hspace{3mm}  0&79  \hspace{3mm} & \hspace{2mm}  1&00 \hspace{2mm} \\
 \hspace{2mm} all       \hspace{2mm}  & C & C & \hspace{3mm}  0&79  \hspace{3mm} & \hspace{2mm}  1&00 \hspace{2mm} \\
 \hspace{2mm} n.n.      \hspace{2mm}  & A & B & \hspace{3mm} -0&57  \hspace{3mm} & \hspace{2mm} -0&50 \hspace{2mm} \\
 \hspace{2mm} n.n.      \hspace{2mm}  & A & C & \hspace{3mm} -0&57  \hspace{3mm} & \hspace{2mm} -0&50 \hspace{2mm} \\
 \hspace{2mm} n.n.      \hspace{2mm}  & B & C & \hspace{3mm} -0&57  \hspace{3mm} & \hspace{2mm} -0&50 \hspace{2mm} \\
 \hspace{2mm} non-n.n.  \hspace{2mm}  & A & B & \hspace{3mm} -0&39  \hspace{3mm} & \hspace{2mm} -0&50 \hspace{2mm} \\
 \hspace{2mm} non-n.n.  \hspace{2mm}  & A & C & \hspace{3mm} -0&39  \hspace{3mm} & \hspace{2mm} -0&50 \hspace{2mm} \\
 \hspace{2mm} non-n.n.  \hspace{2mm}  & B & C & \hspace{3mm} -0&39  \hspace{3mm} & \hspace{2mm} -0&50 \hspace{2mm} \\
\hline\hline
\end{tabular}
\end{table}

\subsection{Spin correlations}
\label{sec:spin_correlations}

With the ground-state CPS wavefunction it is also possible to compute the spin-spin correlation functions.
These are shown in Table \ref{tab:spin_corr} for the case of no external field.
As the spin on the magnetic sites increases from 1/2 to $\infty$, the
resulting ground-state is expected to become increasingly classical. The classical ground-state for corner-sharing triangles
is well-known. Recall that the lattice is tripartite. Then all spins on sub-lattice A (and similarly for B and C) point in the
same direction in the classical ground-state. The relative angle between the spins on sub-lattices A, B, and C is 120 degrees
as is found in the classical ground-state for the Heisenberg triangle. Note that there are an infinity of classical
ground-states, as the plane of the spins for sub-lattices A, B, and C can be rotated continuously.

From our calculated correlation functions, the strongest correlations are naturally within the triangles.
In the classical case, the spins are perfectly rotated from each other by 120 degrees, producing a dot product
$\vec{S}_i \cdot \vec{S}_j / (|\vec{S}_i| |\vec{S}_j|)$ of -0.5 between nearest neighbor spins.
Our CPS ansatz predicts that quantum fluctuations enhance the expectation value of this dot product to -0.57.
In doing so, the parallelity of spins on the same sub-lattice is disrupted.
In the classical case we expect the spins to be perfectly parallel on the same sub-lattice, but quantum
fluctuations reduce the average same sub-lattice dot product from 1.0 (the classical value) to 0.79.
These values are the same for each sub-lattice, showing that at least by this metric, our ansatz preserves
the equivalence of the different sub-lattices.


\begin{figure}[t]
\centering
\includegraphics[width=8.5cm,angle=0]{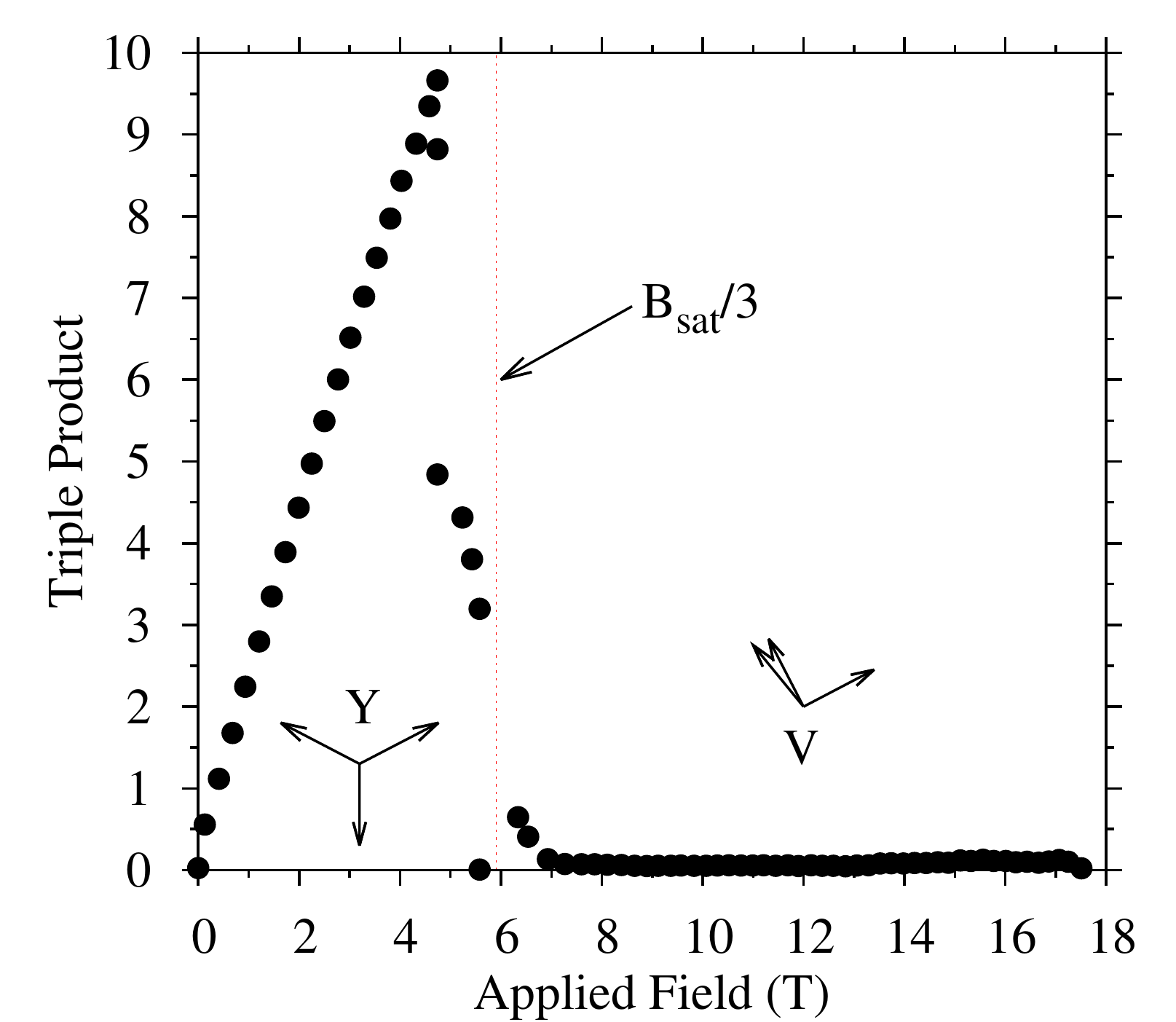}
\caption{The volumes of the parallelepiped defined by the spin triad's three vectors, given by
         the average of the scalar triple products $\vec{S}^i \cdot (\vec{S}^j \times \vec{S}^k)$
         over all triangles with the sites ordered by sublattice ($i\in A,j\in B,k\in C$).
         Note that the individual triangles' triple products all had the same sign and
         that they deviated very little from the average.
         See Section \ref{sec:phase_transition}.}
\label{fig:triple_product}
\end{figure}

\subsection{Remnants of the $B_{\mathrm{sat}}/3$ phase transition}
\label{sec:phase_transition}

\begin{figure}[t]
\centering
\includegraphics[width=8.5cm,angle=0]{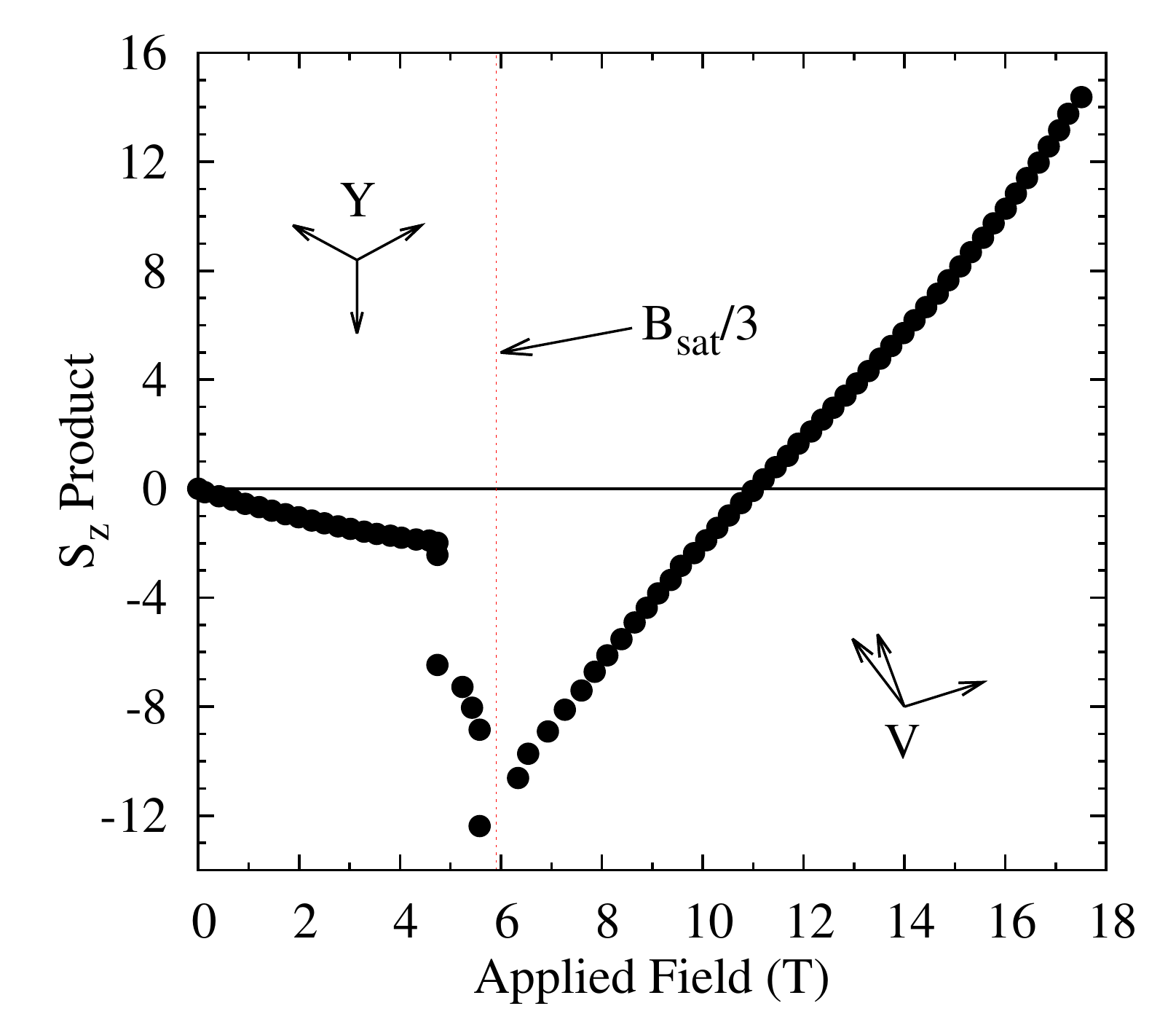}
\caption{The expectation values of the product $S^i_z S^j_z S^k_z$ of the three $z$ components
         of the spin triad's vectors, averaged over all triangles $\{i,j,k\}$.
         Note that the individual triangles' products deviated very little from the average.
         See Section \ref{sec:phase_transition}.}
\label{fig:szszsz}
\end{figure}

\begin{figure*}[t]
\centering
\includegraphics[width=16cm,angle=0]{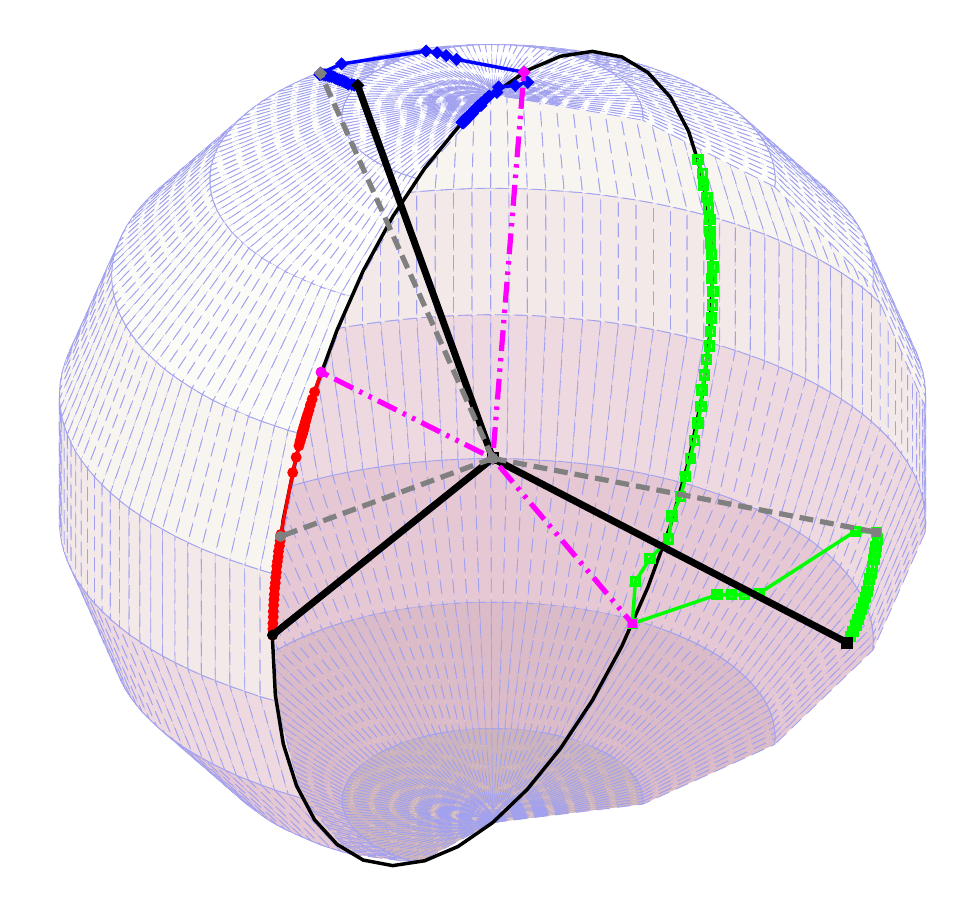}
\caption{Evolution of a triad of classical spin vectors depicting the qualitative changes in our
         wavefunction with increasing applied field.
         The positions of the three vectors' endpoints on the $S^2=35/4$ sphere are given by a red line
         with circles, a green line with squares, and a blue line with diamonds.
         The heavy black lines represent the spin vectors at $B=0$, the dashed grey lines represent
         the spin vectors just before the transition, and the dot-dot-dashed pink lines represent the spin
         vectors just after the transition.
         The thin black circle represents the intersection of the $S^2=35/4$ sphere with the $xz$ plane.
         See Section \ref{sec:phase_transition}.}
\label{fig:sphere}
\end{figure*}

We now seek to provide some qualitative understanding of our
wavefunction at different total $S_z$ values.
In doing so we will show that our wavefunction undergoes a change similar to a phase
transition as the applied field is increased.
In classical corner sharing triangle lattices, phase transitions are known to occur
near $B_{\mathrm{sat}}/3$ between phases in which the spins on the different
sublattices take on ``Y'' or ``V'' shaped configurations \cite{GVOZDIKOVA:2011:phase_diag}.
We must stress, however, that our numerical results do not allow
us to fully distinguish whether our wavefunction's phase transition behavior
is a true property of the $S=5/2$ icosidodecahedron or an artifact of our approximate
ansatz, a point we discuss in some detail below.

To characterize our wavefunction, we will focus on the behavior of the spin triads that make
up each of the icosidodecahedron's triangles, an approach similar to the characterization of
phases in the two dimensional infinite triangular and Kagome lattices.
There, the phases are coplanar and described as either ``Y'' (or umbrella), ``V'', $uud$,
or $uuu$ \cite{GVOZDIKOVA:2011:phase_diag}.
The ``Y'' and ``V'' phases are so named because the shapes of these letters correspond to
how the three spins are arranged in the plane (in the ``V'' case two of the spins are collinear).
We will see that our wavefunction undergoes a sharp change between two states similar to the
classical ``Y'' and ``V'' states, although our ``Y'' state is not coplanar and the spins in our
``V'' state may not be completely collinear (see Figures \ref{fig:triple_product} and
\ref{fig:szszsz} for cartoons).

To probe the character of our wavefunction's spin triads, we have computed two expectation values.
First, we have computed the scalar triple product of the three spin vectors of
each triangle $\vec{S}^i \cdot (\vec{S}^j \times \vec{S}^k)$, which gives the volume of the
parallelepiped that they define.
For coplanar or collinear spins, the triple product will be zero, which should help us
differentiate between these configurations and others, such as a partially folded
umbrella arrangement.
In Figure \ref{fig:triple_product}, we plot the averages (over the twenty triangles) of the
parallelepiped volumes for the ground state wavefunctions at different applied fields.
We see that before $B_{sat}/3$, the volume increases with field strength,
which suggests the state may be a ``folding umbrella'' in which the initially $120^\circ$
rotated spins gradually close towards the $z$ axis.
However, near $B_{sat}/3$, the volume drops abruptly to zero and remains there for all higher
field strengths.
It is tempting to interpret this rapid drop as the remnants of what in classical 2D lattices
would be a phase transition between non-coplanar and coplanar phases, although a recent study
of the classical triangular and Kagome lattices observes only coplanar phases at low
temperatures \cite{GVOZDIKOVA:2011:phase_diag}.
It appears that either quantum effects or errors inherent to our ansatz are stabilizing a
non-coplanar arrangement for applied fields below $B_{sat}/3$.

To further elucidate the qualitative nature of the states before and after $B_{sat}/3$,
we have also computed for each triangle the expectation values of the product of the three
spins' $S_z$ operators, $S^i_z S^j_z S^k_z$.
The average of these quantities over all triangles is shown in Figure \ref{fig:szszsz} for
different applied field strengths.
We see that before $B_{sat}/3$, the $S_zS_zS_z$ expectation values are negative,
indicating that if the state is indeed a folding umbrella in this regime that one of the
three spin vectors lies below the $xy$ plane (this is the only way for three vectors with
a non-negative net $S_z$ to give a negative $S_zS_zS_z$ product).
Thus it appears the state may be a tilted umbrella, in which the axis (or ``handle'') of the
umbrella has been rotated away from vertical in such a way as to place one of the spokes below
the $xy$ plane.
As with the triple product, our $S_z$ product shows an abrupt change near $B_{sat}/3$,
dropping rapidly to a value near the maximum-magnitude negative of $(-5/2)^3$ characteristic
of the $uud$ configuration.
It has been shown \cite{SCHRODER:2005:diff_susc} that the $uud$ state makes a major
contribution to the properties of the classical spin icosidodecahedron near $B_{sat}/3$.
The analogous quantum states play a similar role in the $S=1/2$ icosidodecahedron
\cite{MILA:2008:kagome_sphere}, and the same now appears to be true for $S=5/2$.
Upon increasing the field further, the $S_z$ product rises smoothly to its maximum value at
saturation, indicating that the three coplanar spin vectors are smoothly converting from
an $uud$ type configuration into the $uuu$ configuration.

As a final means to give a qualitative feel for our wavefunction, we have constructed the
classical spin triad that most closely matches the above expectation values.
To do so, we have required that the triple products, $S_z$ products, and total $S_z$
magnetizations match those given by our quantum wavefunction.
In addition, we have arbitrarily restricted one of the three vectors to the $xz$ plane (our
quantum expectation values are all rotationally invariant about the $z$ axis).
Finally, in order to create a unique classical state, we have also required that the $x$
and $y$ components of the classical vectors each add to zero, as is the case for the classical
ground state on a spin triangle \cite{GVOZDIKOVA:2011:phase_diag}.
These requirements give a unique evolution of the classical spin vectors with increasing
applied field strength, which is depicted in Figure \ref{fig:sphere}.
Note that as the state approaches saturation, the quantum expectation values become
incompatible with a classical spin state, and so we have only plotted the spin evolution up to
the point at which compatibility fails.

Unfortunately, we cannot rule out the possibility that the phase change behavior we observe
may be an artifact of our approximate wavefunction.
Our primary concern is that our initial guess is biased towards a particular
coloring of the icosidodecahedron lattice (see Section \ref{sec:wfn_details} and Figure
\ref{fig:coloring}), but this is not the only way to color the lattice and thus 
the initial guess does not possess all the correct symmetries.
While it is possible that the optimization repairs this deficiency, 
it would be preferable
to work with a wavefunction without this handicap.
In future work it may be possible to use as the ansatz a linear combination of CPS states with
an initial guess taken such that each CPS in the combination is biased towards a different
lattice coloring, thus removing fears of a coloring bias in the overall ansatz.
While our computer implementation is not currently capable of optimizing such an ansatz,
we do not foresee any fundamental barriers to executing such an optimization in the future.

\section{Conclusions}
\label{sec:conclusions}

In this work we demonstrated that the correlator product state, a very simple ansatz designed for the treatment
of strongly correlated spins, can successfully be used to model the quantum states of complex molecular magnets
such as the Fe$_{30}$ Keplerate system. The size of this system lies far outside the range of exact diagonalization.
Our calculated variational energies are significantly lower than those previously obtained with the density matrix renormalization
group and produce a fit to the rotational band model that is almost identical to that derived from experimental
magnetization data.
Furthermore, unlike the rotational band model, our ansatz is capable of predicting anomalies in the differential susceptibility
and heat capacity that are observed in frustrated magnetic systems near 1/3 of the saturation field.
We have also analyzed  a number of correlation functions among the spins, showing how the quantum state deviates
from classical behavior.
Finally, we have shown how as a function of magnetic field, the quantum state appears to undergo a change
reminiscent of phase transitions seen in classical 2D corner sharing triangular lattices.
 
In future research, more work is needed to  clarify a number of aspects of this study.
While our variational energies are superior to previous theoretical treatments, neutron scattering data suggests
there is still ample room for improvement, especially for small applied fields, as our predicted singlet-triplet
gap is too small. In addition, it is not clear that the optimization of our ansatz fully preserves
all symmetries of the icosidodecahedron, which makes definitive conclusions regarding the observed phase transition
behavior difficult. 
To address these shortcomings, we have suggested that an ansatz consisting of a specially crafted linear combination
of correlator product states be employed.
The other obvious omission is a treatment of excited states, given that the presence of low-lying
excitations is a key feature of frustrated spin systems.
Generalizing the methodology to model excited states within the Monte Carlo framework
 is not as straightforward as the generalization to linear
combinations, but the critical importance of low-lying excitations makes it a highly desirable goal.

The CPS family of states can naturally be applied to other magnetic systems as well as to more general
non-spin electronic systems \cite{NEUSCAMMAN:2012:fast_sr}.
In the latter case, it is more advantageous to combine the correlators with a fermionic reference function.
Correlators used in this way are formally the same as the Jastrow factors long studied in electronic structure.
Jastrow factors are usually employed to model ``weak'' correlations associated with the electron-electron cusp,
while correlators have proven effective at introducing strong correlations in the Hubbard model and some
molecular systems. In the present study, we have shown that even the very complex correlations arising from magnetic frustration can
be described effectively using correlators.
Taken together, these findings motivate the use of correlator product states as a means to describe
both weak and strong electron correlations simultaneously, a prospect that is under active investigation.


\section{Acknowledgments}
\label{sec:acknowledgments}

\newlength{\myLength}
\setlength{\myLength}{\parindent}
\noindent
\begin{minipage}[h]{\linewidth}
\setlength{\parindent}{\myLength}
This work was supported by NSF EAGER CHE-1004603, 
the David and Lucile Packard Foundation, and by the 
Miller Institute for Basic Research in Science.
\end{minipage}
\vspace{1mm}
\setlength{\parindent}{\myLength}

\bibliographystyle{aip}
\bibliography{cps_mo72fe30.bib}

\end{document}